\documentclass[12pt]{article}
\usepackage{natbib}
\usepackage{epsfig}
\begin{document}
%
\begin{center}
{\bf\Large Propagation of Errors for Matrix Inversion}\\
\vspace{1cm}
{M. Lefebvre\footnote{Department of Physics and Astronomy, P.O. Box 3055
 University of Victoria, Victoria, BC, Canada, V8W-3P6},
R.K. Keeler$^1$,
R. Sobie$^{1,}$\footnote{Supported by the Institute of Particle Physics of Canada},
J. White$^{1,}$\footnote{Present address: Department of Physics, 
Carleton University,  \\ Ottawa, Ontario, Canada}}
%
%
%
%
%
\end{center}

\begin{abstract}
A formula  is given for the propagation of errors during 
matrix inversion. An explicit calculation for a $2\times2$
matrix using both the formula and a Monte Carlo calculation are
compared. A prescription is given to determine when a matrix with
uncertain elements is sufficiently  nonsingular for the calculation
of the covariances  of the inverted matrix elements to be reliable. \\
\end{abstract}
\section{Introduction}
There  are many  problems that involve  solving a  set of simultaneous
linear equations. In many instances, the coefficients of the variables
are uncertain and these uncertainties need to be taken into account in
the  final solution.  The results  presented in this paper are general
and  can be  applied to any  problem involving  the solution of linear
equations. Moreover the full covariance matrix  is calculated. The use
of  the covariance formula is   illustrated  using an example where  a
number of  branching ratios of a given particle were simultaneously 
measured.

A   recent paper reporting  results on  the  decay  of  the tau lepton
\cite{ref_OPAL} used a matrix inversion technique to solve for several
branching ratios simultaneously.  In that paper,  the treatment of the
statistical  uncertainties assumed  only  diagonal  errors. This paper
develops a formula for the covariance of  the inverse matrix elements.
The    treatment   follows    the   propagation  of  errors  formalism
\cite{ref_PDG}  for small errors.  A full  treatment of the covariance
matrix is     necessary  because the   off-diagonal errors     can  be
significant.

The fraction of time  a particle decays into  a given final state
is defined  to  be the  branching ratio.  The  branching  ratios of  a
particle decaying into   a set of $m$  possible  final states may   be
determined by finding  $m$ selection criteria  that have an efficiency
for  selecting the desired decay channel.   The selection criteria are
usually not fully efficient so a set of linear equations results,
\begin{eqnarray}
   \epsilon_{11}B_1  +  \epsilon_{12}B_2  +  
        \epsilon_{13}B_3  +\cdots+  \epsilon_{1m}B_m &
               =  & f_1    \nonumber  \\
   \epsilon_{21}B_1  +  \epsilon_{22}B_2  + 
        \epsilon_{23}B_3  +\cdots+  \epsilon_{2m}B_m &
               =  & f_2     \nonumber  \\
   \epsilon_{31}B_1  +  \epsilon_{32}B_3  +
        \epsilon_{33}B_3  +\cdots+  \epsilon_{3m}B_m &
               =  & f_3                        \label{ref_lin_eqtns} \\
                & \vdots &      \nonumber  \\
   \epsilon_{m1}B_1  +  \epsilon_{m2}B_2  + 
              \epsilon_{m3}B_3  +\cdots+  \epsilon_{mm}B_m &
                     =  & f_m      \nonumber
\end{eqnarray}
where $B_j$ is the $j^{\mathrm{th}}$ unknown branching ratio, $f_i$ is
the fraction of events chosen  by the $i^{\mathrm{th}}$ selection from
a  measured sample with  an  efficiency $\epsilon_{ij}$. The fractions
are usually corrected for background using Monte Carlo estimates.

The efficiency is typically calculated by applying the same selections
to a Monte Carlo sample of events. Thus, the efficiency is given by
\begin{equation}
\epsilon_{ij} =
\frac{n^{\mathrm{MC}}_{ij}}{N^{\mathrm{MC}}_{j}}\:,
                                                      \label {eqteff}
\end{equation}
where $n^{\mathrm{MC}}_{ij}$ is the number  of events of decay channel
$j$ selected by selection $i$  and $N^{\mathrm{MC}}_{j}$ is the  total
number  events of type $j$  in the Monte  Carlo sample. As can be seen
from    equation~\ref{eqteff},  the  elements of  $\epsilon_{ij}$  are
positive definite and less  than or equal to  one.  An uncertainty for
each efficiency matrix element can be  determined from the Monte Carlo
statistics.

The simultaneous equations given by equation~\ref{ref_lin_eqtns} 
can be written in matrix form,
\begin{equation}
           \epsilon B = f\: .    \label{eqtn_sum_mode}
\end{equation}
The vector of branching ratios can  be solved from the matrix
equation,
\begin{equation}
B = \epsilon^{-1}f\:,
\end{equation}
where  $\epsilon^{-1}$  is the inverse of   the efficiency matrix. The
matrix  $\epsilon$ must  be nonsingular  in  order for the inverse  to
exist  or equivalently the   determinant must be  nonzero. A  singular
$\epsilon$ signals an ill defined set of selections or event types.

The uncertainty of the branching ratios, $B$,  can be written in terms
of    the uncertainties  on the  elements    of  the  matrices $f$ and
$\epsilon^{-1}$.  The errors for $f$ are determined  from the data and
the estimates of the background.  The errors on $\epsilon^{-1}$ can be
calculated  in  terms   of   the  covariances  of    the  elements  of
$\epsilon$. The  error  formulae   for $\epsilon^{-1}$  used   in  the
literature are described in section~2. A more general treatment of the
errors taking into account the  full covariance matrix is presented in
section~3 whilst the detailed derivations are given in Appendix~A.

An analytical  example  based   on matrices  of   order $2\times2$  is
developed   in  Appendix~B and  a  Monte  Carlo   study  based on  the
$2\times2$ case is discussed  in  sections~4~and~5.  The Monte   Carlo
studies will illustrate  that the formula  is only  realistic when the
efficiency  matrix is sufficiently far away  from being  singular. The
calculation of the covariances becomes more reliable as  the  ratio of
the value of the determinant of the efficiency matrix, to the value of 
the uncertainty on  the determinant  becomes larger. A formula for the
uncertainty of   a determinant is derived  in   Appendix~C.  The paper
concludes with some general observations.

\section{Calculation of Errors}
Errors  are often estimated by ignoring  the  off-diagonal elements of
the   covariance  matrix.  This is    the   correct procedure  if  the
quantities  are independent   of each other.   A   calculation  of the
uncertainty on the  branching  ratios, assuming  the elements of   the
inverted matrix  are  statistically independent of  each other, yields
\footnote{Square brackets have been  used to separate the superscripts
and subscripts used   to identify the quantity   of interest from  any
indices and mathematical operations.},
\begin{equation}                 
    [\sigma_{B}]^{2}_{i} = [\epsilon^{-1}]^{2}_{ij}[\sigma_{f}]^{2}_{j}
       + [\sigma_{{\epsilon}^{-1}}]^{2}_{ij}[f]^{2}_{j} \:,       
                                                       \label{eqtbasic}
\end{equation}  
where  the  uncertainties   are denoted by  $\sigma$ and   a  sum over
repeated indices is assumed unless otherwise noted.  The errors on the
elements       of     the    inverted         efficiency       matrix,
$\sigma_{{\epsilon}^{-1}}$, are  calculated from  the known  errors on
the efficiency matrix.

The uncertainties on the  inverse efficiencies, have  been determined
(see  for  example reference~\cite{ref_OPAL}) by  differentiating the
matrix equation $\epsilon \epsilon^{-1} =  I$ and then applying a
matrix analogy to the error propagation formula to yield,
\begin{equation}
[\sigma_{{\epsilon}^{-1}}]_{ij}^2 = 
\mid [\epsilon^{-1}]_{im}[\sigma_\epsilon]_{mn}
[\epsilon^{-1}]_{nj} \mid ^{2}\:,                 \label{eqtsimple}
\end{equation}
which  has  the   same  form as  a  transformation    to a  new  basis
\cite{LYONS}.   Equation~\ref{eqtsimple} is  a reasonable approximation
but is  not correct.  It also   neglects any correlations  between the
elements of the inverse matrix,  $\epsilon^{-1}$, which
can be significant, as shown in the next section.

\section{The Covariance of the Elements of \protect $\epsilon^{-1}$}
Matrix inversion  is a nonlinear  operation. It is always  possible to
write the inverse  of  a matrix in  terms of  the matrix  of cofactors
divided by the determinant~\cite{ref_mathbook}. One sees explicitly in
Appendix~C that each element of an inverse matrix  has elements of the
original matrix in   common.  Therefore the  inverse   matrix elements
clearly are correlated.

Consider the  $m^2$ matrix elements $\epsilon_{ij}$ with uncertainties
given    in the   most  general   case    by  the  $m^4$   covariances
$\mathrm{\mathrm{cov}}(\epsilon_{\alpha\beta},\epsilon_{ab})$.     The
inverse    matrix  elements $\epsilon^{-1}_{ij}$,    in  general, have
covariances
$\mathrm{\mathrm{cov}}(\epsilon^{-1}_{\alpha\beta},\epsilon^{-1}_{ab})$,
which can be written as,
\begin{equation}
\mathrm{cov}
(\epsilon^{-1}_{\alpha \beta}, \epsilon^{-1}_{a b}) =
\epsilon^{-1}_{\alpha i} \epsilon^{-1}_{j \beta} 
\epsilon^{-1}_{a k} \epsilon^{-1}_{l b} \mathrm{cov} (\epsilon_{i j}, 
\epsilon_{k l}) \:.
\label{eqt_final}
\end{equation}
The full  derivation of this  equation  is  given in  Appendix A  (see
equation~\ref{eqtfinal}).  The  usual     case where   there are    no
correlations  between  the  elements   of the efficiency  matrix  (see
equations~\ref{eqtAdiag}, \ref{eqtdiagonal} and \ref{eqtfull_cov_diag})
is given by
\begin{equation}
\mathrm{cov}
\left(
\epsilon_{ij},\epsilon_{kl}
\right) = [\sigma_{\epsilon}]^{2}_{ij} 
\delta_{ik}\delta_{jl} \: \:\: \mathrm{ (no \: \: summation) }.
\label{eqt_diag}
\end{equation}
Hence the full set of covariances of $\epsilon^{-1}$ are given by
\begin{equation}
\mathrm{cov}(\epsilon^{-1}_{\alpha\beta},\epsilon^{-1}_{a b}) =
([\epsilon^{-1}]_{\alpha i}[\epsilon^{-1}]_{a i})
[\sigma_{\epsilon}]^2_{ij}
([\epsilon^{-1}]_{j \beta}[\epsilon^{-1}]_{j b}) \:,
\label{eqtfulldiagcov}
\end{equation}
where  there is no  sum in this  case over repeated indices inside the
parentheses.    The variance of  an element  of the inverse efficiency
matrix can be written as,
\begin{equation}
[\sigma_{\epsilon^{-1}}]_{\alpha\beta}^2  
        \equiv  
\mathrm{cov}(\epsilon^{-1}_{\alpha\beta},\epsilon^{-1}_{\alpha\beta}) 
  =  [\epsilon^{-1}]_{\alpha i}^2
[\sigma_{\epsilon}]^2_{ij}              \label{eqt_diag_var}
[\epsilon^{-1}]_{j \beta}^2 \:.
\end{equation}
This  equation   is      exact    and  replaces      the   approximate
equation~\ref{eqtsimple}.     Note        that     each     term    of
equation~\ref{eqt_diag_var} is  squared before  making the sum whereas
in equation~\ref{eqtsimple} the sum is done first.

The complete expression for the uncertainties on the branching ratios
can then be calculated from,
\begin{equation}
\mathrm{cov}(B_{i},B_{j})=f_{\alpha}f_{\beta}
\mathrm{cov}(\epsilon^{-1}_{i\alpha},              \label{eqtwholething}
\epsilon^{-1}_{j\beta}) +
\epsilon^{-1}_{ik}\epsilon^{-1}_{jl}\mathrm{cov}(f_{k},f_{l}) \:,
\end{equation}
where the measured fractions of events are often uncorrelated,
\begin{equation}
\mathrm{cov}(f_k,f_l) 
   = [\sigma_f]^2_k \delta_{kl}  \: \:\: \mathrm{ (no \: \: summation) }.
\end{equation}
Equation~\ref{eqtwholething} is a generalization of 
equation~\ref{eqtbasic}, including all of the covariances.

\section{A Monte Carlo Study - Part 1}
The  properties of   equation~\ref{eqtfulldiagcov}  have  been studied
using a Monte Carlo  simulation  of a  $2\times2$  matrix.  Relatively
small errors have  been used to satisfy  the small error approximation
used  to  derive the  error  propagation  formula~\cite{ref_PDG}.  The
results  of the Monte Carlo  calculations directly can  be compared to
the analytic formulae developed in Appendix~B.

Given an   initial efficiency matrix $\epsilon$,   and the variance on
each   element, $[\sigma_{\epsilon}]^{2}_{ij}$,   the  $m^{th}$ random
instance of the matrix is generated by
\begin{equation}
[\epsilon_m]_{ij} = \epsilon_{ij} + 
    \Gamma_{m} \cdot [\sigma_{\epsilon}]_{ij} \:,
\end{equation}
where $\Gamma_{m}$ is a normally distributed pseudorandom deviate with
a  mean of zero and a  standard deviation of  one. A set of $N$ matrices
are created, are then inverted.  The covariances are calculated from
\begin{equation}
\mathrm{cov}(\epsilon^{-1}_{\alpha \beta}, \epsilon^{-1}_{a b})
= \langle \epsilon^{-1}_{\alpha \beta}~ \epsilon^{-1}_{a b}\rangle
 - \langle  \epsilon^{-1}_{\alpha \beta}\rangle
\langle \epsilon^{-1}_{a b} \rangle \:,
\label{eqtmvcov}
\end{equation}
where $\langle~~ \rangle$ is the mean of the  quantity enclosed in the
brackets.

A sample of $N=10,000$ instances of the matrix 
$\epsilon \pm \sigma_{\epsilon}$ where,
\begin{equation}
  \epsilon=
    \left(    \begin{array}{cc}
           0.700 & 0.200 \\
           0.400 & 0.600 
          \end{array}    \right) 
  \pm
     \left(    \begin{array}{cc}
           0.007 & 0.002 \\
           0.004 & 0.006 
          \end{array}    \right) \:,      
\label{eqt_ns_matrix}
\end{equation}
were generated with each element of $\epsilon$ normally distributed 
about the central
values with the one percent errors, $\sigma_{\epsilon}$, indicated.
The inverse of the matrix $\epsilon$ is given by
 \begin{equation}
  \epsilon^{-1} =
    \left(    \begin{array}{cc}
           1.765 & -0.588\\
           -1.177 & 2.059 
          \end{array}    \right) \:,
\label{eqt_ns_inverse}
\end{equation}
and the determinant, $\det \epsilon = 0.340$~ The covariances can
be calculated using only the inverse matrix, $\epsilon^{-1}$, and the
error matrix, $\sigma_{\epsilon}$. The numerical values 
for the 16 covariances of $\epsilon^{-1}$, calculated from the Monte Carlo 
and analytically using equation~\ref{eqtfulldiagcov}, are given in 
Table~\ref{tblsingularcov1}. The symmetry of the covariance operation
implies there should be 10 unique numbers as is apparent from
the table.
%
%
\begin{table}[ht]
\begin{center}
  \begin{tabular}{|rr|rr|rr|} \hline
    \multicolumn{6}{|c|} {Covariance Matrix $\epsilon^{-1}$} \\ \hline
    \multicolumn{2}{|c} {Monte Carlo $\times(10^{-4})$ } & 
\multicolumn{2}{|c|}{Analytic $\times(10^{-4})$} &
                       \multicolumn{2}{c|}{Fractional Difference} \\ \hline
    \multicolumn{6}{|c|} 
          {$\mathrm{cov}(\epsilon_{11}^{-1},\epsilon_{ij}^{-1})$ } \\ \hline
$5.231$  & $-2.312$ &  $5.269$  & $-2.245$ &  $-0.007$ & $0.029$  \\
$-4.604$ & $2.496$ & $-4.490$ & $2.514$ & $0.025$  & $-0.007$ \\ \hline
    \multicolumn{6}{|c|}
 {$\mathrm{cov}(\epsilon_{12}^{-1},\epsilon_{ij}^{-1})$ } \\ \hline
$-2.312$ & $1.643$ & $-2.245$ & $1.603$ & $0.029$  & $0.024$  \\
$2.567$ & $-2.642$ & $2.514$ & $-2.619$ & $0.020$ & $0.009$  \\  \hline
    \multicolumn{6}{|c|}
 {$\mathrm{cov}(\epsilon_{21}^{-1},\epsilon_{ij}^{-1})$ } \\ \hline
$-4.604$ & $2.567$ & $-4.490$ & $2.514$ & $0.025$  & $0.020$  \\
$6.511$ & $-5.229$ & $6.413$ & $-5.238$ & $0.025$ & $0.020$  \\ \hline
    \multicolumn{6}{|c|}
 {$\mathrm{cov}(\epsilon_{22}^{-1},\epsilon_{ij}^{-1})$ } \\ \hline
$2.496$ & $-2.642$ & $2.514$ & $-2.619$ & $-0.007$& $0.009$ \\
$-5.229$ & $6.971$ & $-5.238$ & $7.172$ & $-0.002$ & $-0.029$  \\ \hline
  \end{tabular}  
\caption{ The covariance matrix for each element of $\epsilon^{-1}$
calculated by the Monte Carlo and calculated using the analytic
formula. The last column is the fractional difference between the
Monte Carlo and analytic calculation.}
  \label{tblsingularcov1}
\end{center}
\end{table}

\begin{figure}
\begin{center}
  \epsfig{file=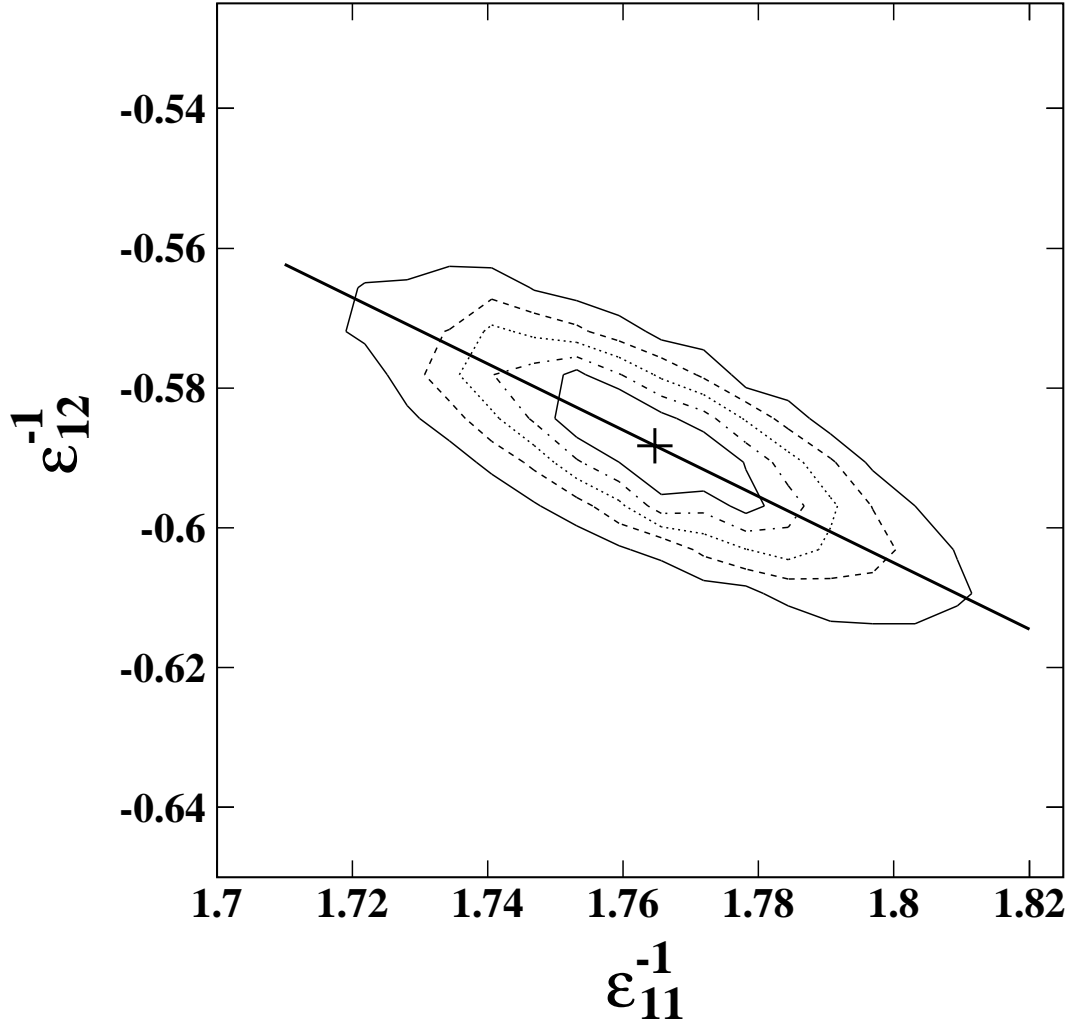,width=15 cm}
     \caption{The 10,000 elements $\epsilon^{-1}_{12}$
and $\epsilon^{-1}_{11}$  generated in the Monte Carlo simulation
form a two dimensional distribution shown here as a contour plot.
The contours are at the level of 50, 100, 150, 200 and 250 counts
per bin (each bin is $6.25 \times 10^{-3} \times 6.25 \times 10^{-3}$). 
The correlation between the two inverse matrix elements
is clear. The data are centred around the exact calculation of the inverse,
shown by the large ``+'' sign.
The slope of the line is the theoretical calculation of the correlation
coefficient. It is in good agreement with the slope of the axis of the
approximately elliptical contours.}
\label{figcorr2d}
\end{center}
\end{figure}

Figure~\ref{figcorr2d} shows a contour plot of the 
element $\epsilon^{-1}_{12}$ versus $\epsilon^{-1}_{11}$ generated
from the Monte Carlo. The contours are approximate lines of
constant probability density.  
The entries are distributed around the values
of the calculated inverse matrix elements and are clearly correlated.
The slope of the error ellipses has been calculated by substituting 
the analytical values of the covariances from Table~\ref{tblsingularcov1} 
into a formula derived from reference~\cite{ref_PDG}.
%
The calculated slope is in good agreement with the slope of the
major axis of the error ellipses. 
%
%

\begin{figure}
\begin{center}
  \epsfig{file=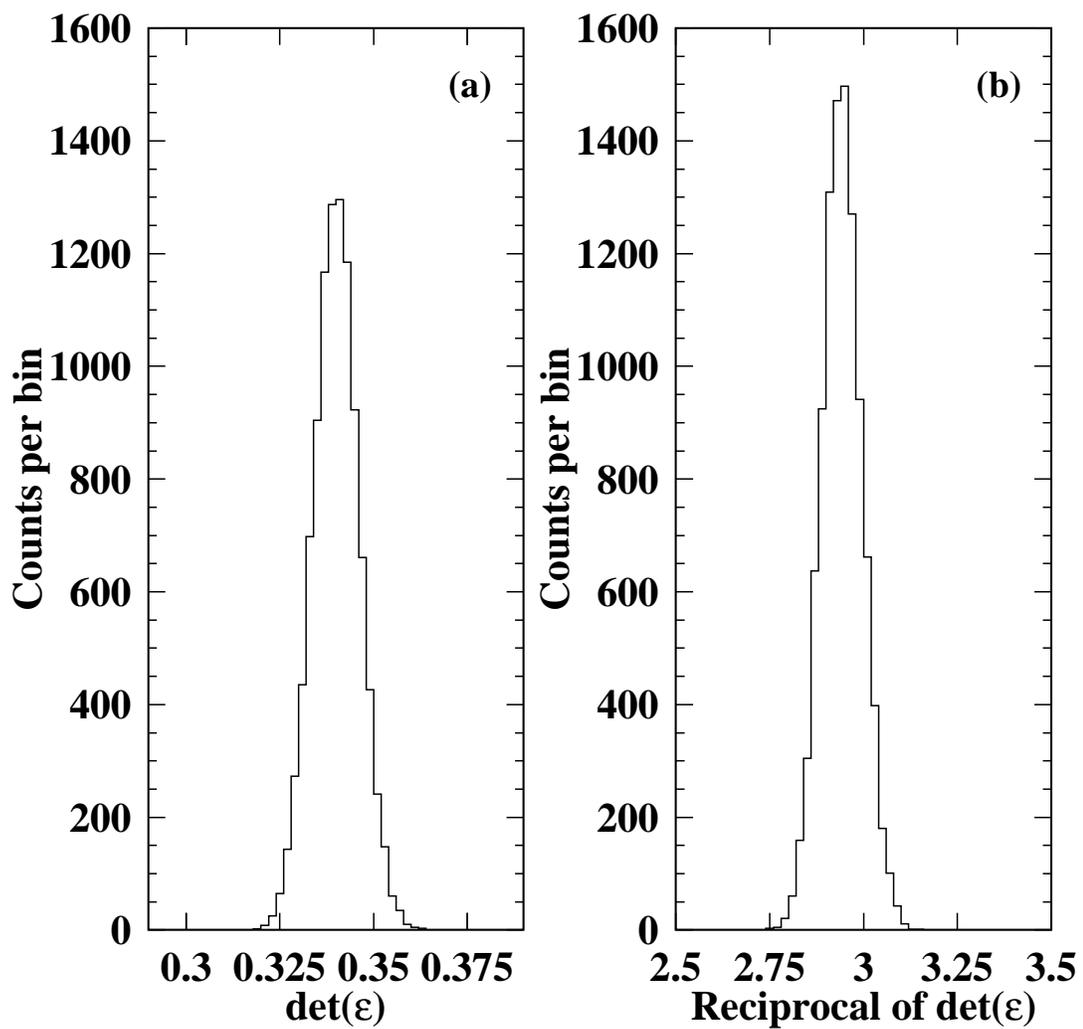,width=15cm}
     \caption{The histogram labelled (a) shows the determinant of the 
matrix $\epsilon$ described
in the text. The values of the determinant are clearly not near zero
and consequently the histogram of the reciprocal of the determinant 
shown in (b) is  well behaved. }
\label{fignonsingdet}
\end{center}
\end{figure}

Recall that the efficiency matrix must be nonsingular in order to invert 
it. In the Monte Carlo
calculation the individual elements of $\epsilon$ are varied and hence,
if the variations are large enough, it is possible for the determinant to 
become zero. Figure~\ref{fignonsingdet}a shows a histogram of the 
Monte Carlo calculation
of the determinant. The mean value is $0.340$ as expected with a root
mean square deviation of $0.006$~. The shape of the distribution is 
normal, see figure~\ref{fignonsingdet}, also as expected given
the determinant is the sum of products of normally distributed quantities.
In this particular case the mean value of the determinant is 
$56.7$ standard deviations from zero. Increasing the errors or 
reducing the determinant will both produce a higher likelihood of the 
determinant fluctuating nearer to zero.
%
%

As noted earlier the inverse matrix elements all have a common factor
of $1/|\epsilon|$. Therefore
values of the determinant near zero produce very large values for the inverse
matrix elements. Moreover, the reciprocal of a normal distribution is 
not normally distributed but is instead asymmetric with a tail towards 
large values. In figure~\ref{fignonsingdet}b one sees that any tails
are insignificant for determinants many standard deviations away from zero.

\begin{figure}
\begin{center}
  \epsfig{file=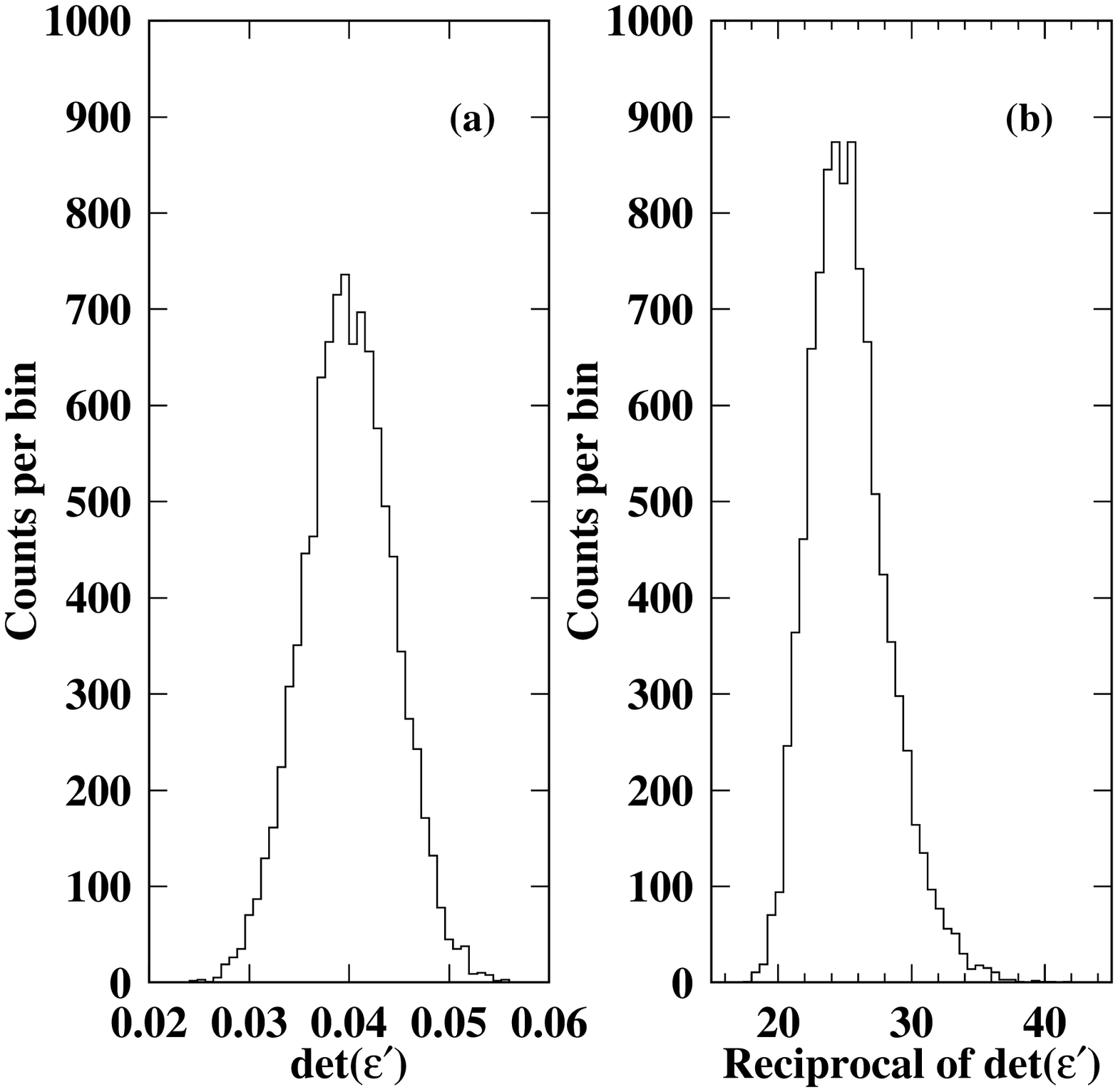,width=15cm}
     \caption{The histogram labelled (a) shows the determinant of 
the matrix $\epsilon$ described
in the text. The values of the determinant approach zero and consequently
an asymmetry is observed in the histogram, labelled (b), of 
the reciprocal distribution. }
\label{figsingdet}
\end{center}
\end{figure}

Figure~\ref{figsingdet} shows an identical analysis for the matrix,
\begin{equation}
  \epsilon^{\prime} =
    \left(    \begin{array}{cc}
           0.400 & 0.500 \\
           0.400 & 0.600 
          \end{array}    \right) 
  \pm
     \left(    \begin{array}{cc}
           0.004 & 0.005 \\
           0.004 & 0.006 
          \end{array}    \right)\:.     
\label{eqt_singular_matrix}
\end{equation}
In this case $\det \epsilon^{\prime}=0.040$ and its root
mean square deviation, due to the one percent
errors, is $0.004$~. Now the distribution of the reciprocal determinants,
seen in Fugure~\ref{figsingdet}b, is clearly skewed. 
%
%
%
Table~\ref{tblsingularcov2} lists the 
values of the covariance matrix calculated for each
element of the inverse matrix, $\epsilon^{ \prime -1}$, using the Monte Carlo
method, equation~ \ref{eqtmvcov} and the analytic expression, 
equation~\ref{eqtfulldiagcov}. 

In Table~\ref{tblsingularcov1} the percent difference 
between the Monte Carlo and the
analytic formula has a mean value of approximately one percent averaged
over the 10 independent values of the covariance. This is consistent with
the precision expected for a 10,000 event Monte Carlo and demonstrates
that the analytic expression reproduces the covariances  well
for the far from singular case. However, the 
same mean for the values in Table~\ref{tblsingularcov2} 
is eleven percent. Moreover, the
magnitude of the Monte Carlo value is larger than the analytic value for 
each covariance. The analytic expression becomes less accurate
and no longer appropriate to use as the matrix becomes closer to being
singular.
%
%
\begin{table}[!ht]
\begin{center}
  \begin{tabular}{|rr|rr|rr|} \hline
    \multicolumn{6}{|c|} {Covariance Matrix $\epsilon^{\prime -1}$} \\ \hline
    \multicolumn{2}{|c} {Monte Carlo $\times(10^{-4})$} & 
\multicolumn{2}{|c|}{Analytic $\times(10^{-4})$} &
                       \multicolumn{2}{c|}{Fractional Difference} \\ \hline
    \multicolumn{6}{|c|}
{$\mathrm{cov}(\epsilon^{\prime -1}_{11},\epsilon^{\prime -1}_{ij})$} \\ \hline
    $2.812$ & $-2.554$ & $2.498$ & $-2.269$ &  $0.112$ & $0.112$  \\
    $-2.040$ & $1.855$ & $-1.815$ & $1.650$ & $0.110$ & $0.110$ \\ \hline
    \multicolumn{6}{|c|}
{$\mathrm{cov}(\epsilon^{\prime -1}_{12},\epsilon^{\prime -1}_{ij})$} \\ \hline
    $-2.554$ & $2.337$ &  $-2.269$ & $2.078$ & $0.112$ & $0.111$  \\
    $1.855$ & $-1.699$ & $1.650$ & $-1.513$ & $0.110$ & $0.110$  \\  \hline
    \multicolumn{6}{|c|}
{$\mathrm{cov}(\epsilon^{\prime -1}_{21},\epsilon^{\prime -1}_{ij})$} \\ \hline
    $-2.040$ & $1.855$ & $-1.815$ & $1.650$ & $0.110$ & $0.110$  \\
    $1.492$ & $-1.358$ & $1.330$ & $-1.210$ & $0.109$ & $0.109$  \\ \hline
    \multicolumn{6}{|c|}
{$\mathrm{cov}(\epsilon^{\prime -1}_{22},\epsilon^{\prime -1}_{ij})$} \\ \hline
    $1.855$ & $-1.699$ & $1.650$ & $-1.513$ & $0.110$ & $0.111$ \\
    $-1.358$ & $1.244$ & $-1.210$ & $1.110$ & $0.109$ & $0.108$  \\ \hline
  \end{tabular}
  \caption{ The covariance matrix for each element of $\epsilon^{\prime-1}$
calculated by the Monte Carlo and calculated using the analytic
formula. The last column is the fractional difference between the
Monte Carlo and analytic calculation.}
  \label{tblsingularcov2}
\end{center}
\end{table}

\section{ A Monte Carlo Study - Part 2}
In order to explore the range of uncertainty on how well the analytic
expression models the covariances, the Monte Carlo simulation described in
Part~1 was modified. Instead of specifying a particular $2\times2$ matrix
for analysis, 5000 $2\times2$ matrices were generated by choosing the elements
of each matrix from a uniform distribution in the interval $[0,1]$~.
Each of the randomly generated matrices was then analyzed using exactly
the same procedure previously discussed. One percent errors were used
as before.

None of the matrix inversions actually failed. However, many cases
involved determinants close to zero.
Figure~\ref{figdiff}(a) shows the fractional 
difference between the Monte Carlo  
and the analytic calculation for 
$\mathrm{cov}(\epsilon^{-1}_{11},\epsilon^{-1}_{21})$,
\begin{equation}
\frac{ \mathrm{cov}(\epsilon^{-1}_{11},\epsilon^{-1}_{21})_
{\mbox{\scriptsize{MC}}} -
 \mathrm{cov}(\epsilon^{-1}_{11},\epsilon^{-1}_{21})_
{\mbox{\scriptsize{Analytic}}}}
{ \mathrm{cov}(\epsilon^{-1}_{11},\epsilon^{-1}_{21})_
{\mbox{\scriptsize{MC}}}}, 
\end{equation}
plotted against
the determinant of $\epsilon$ divided by its 
uncertainty, $\det \epsilon / \sigma_{|\epsilon|}$.
The number of sigmas
the determinant may be from zero can be shown to be $\pm 70.7$ for
one percent errors. The fractional differences between the Monte Carlo
and analytic evaluations are distributed around approximately zero
for large numbers of sigma. However, near 10 sigmas the Monte Carlo
calculated covariance becomes very large and the distribution
becomes centred around one. Under the scatter plot is figure~\ref{figdiff}(b).
This is a plot of the mean value of the fractional difference plotted
against the number of sigmas the determinant is away from zero.
The error bars are the root mean square deviation of the distribution.
The means for $-10. < \det \epsilon / \sigma_{|\epsilon|} < 10. $ 
are not plotted.
The largest deviation, for the bin 10--20, is less than four percent.
The sigmas are approximately two percent. Hence, the analytic formula
does a very good job of estimating the covariance as long as the
determinant of the matrix is at least 10 sigmas from zero. A simple
expression to calculate the error on a determinant is given in Appendix~C.
%
%
%
\setlength{\unitlength}{1mm}
%
\begin{figure}  
   \begin{picture}(160,160)(0,0)
       \put(5,45){\epsfig{file=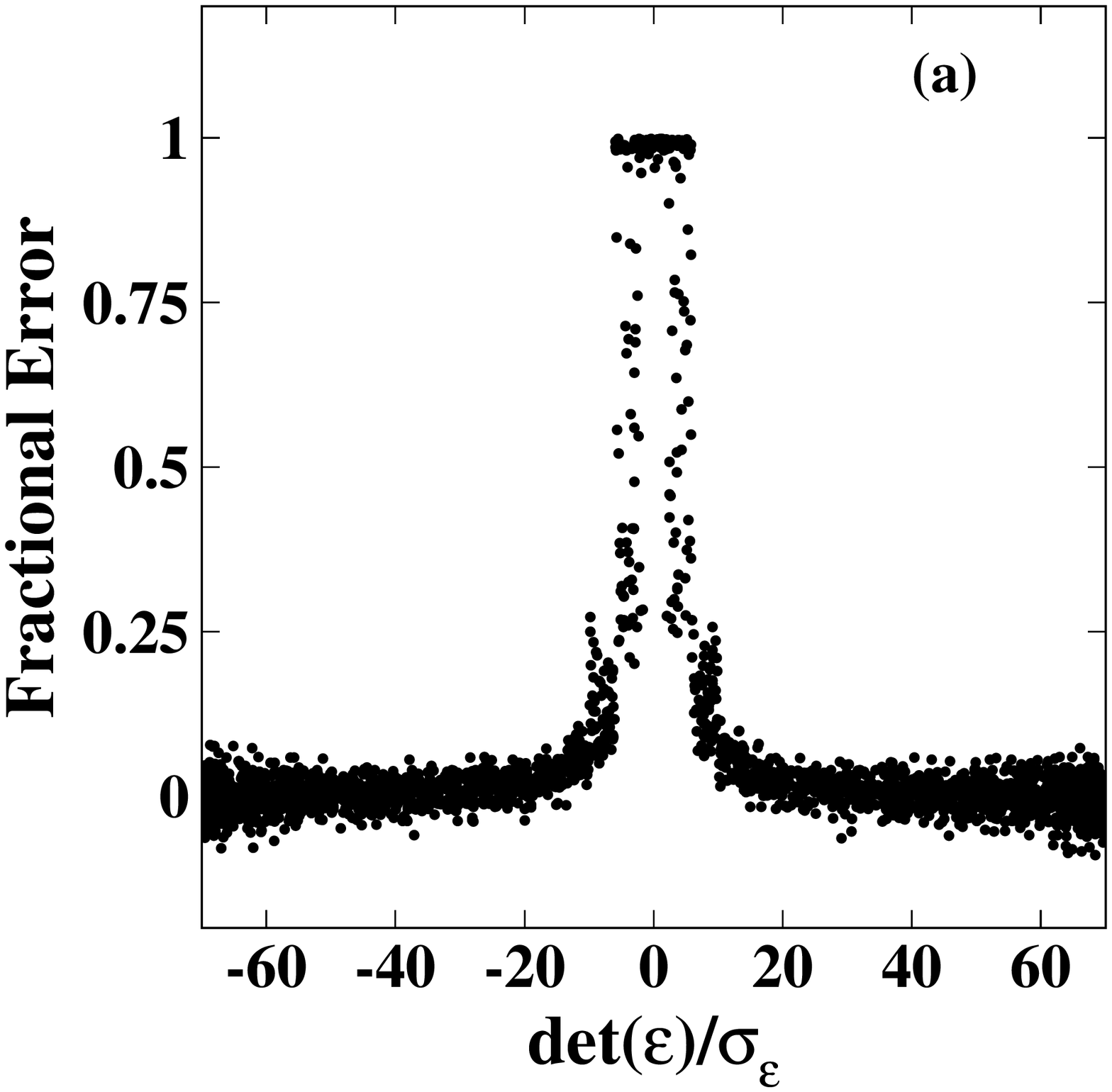,width=11cm}}
       \put(5,0){\epsfig{file=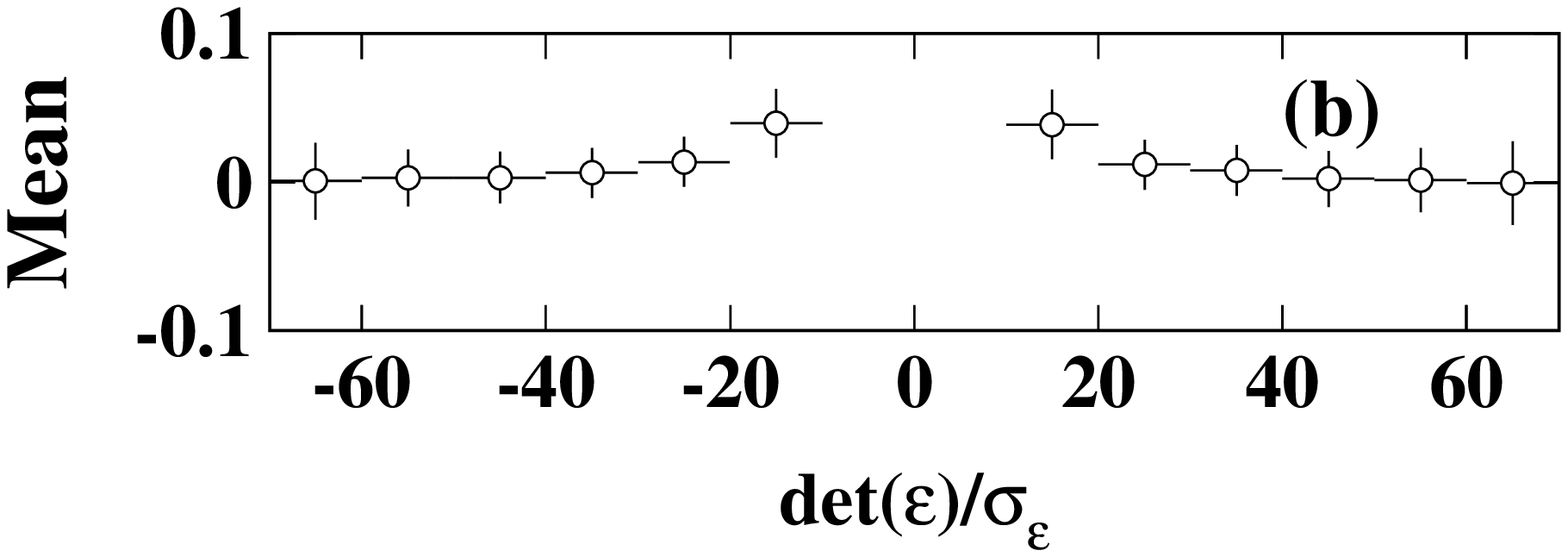,width=11cm,height=4.5cm}}
   \end{picture}
   \caption{The fractional difference between the Monte Carlo
and analytic calculations as a function of
$\det \epsilon / \sigma_{|\epsilon|}$ for 
$\mathrm{cov}(\epsilon^{-1}_{11},\epsilon^{-1}_{21})$ 
is plotted in part (a). The
fractional difference is normalized by the Monte Carlo value.
In part (b) the mean and sigma are plotted for slices of 
distribution (a). There are 14 slices of 
$\det \epsilon / \sigma_{|\epsilon|}$ between -70 and 70. The
two centre slices are not plotted.}
\label{figdiff}
\end{figure}

\section{Conclusions}
The   covariances  of   inverse   matrix  elements   are  nonzero   in
general. Therefore, the propagation of errors for formulae that depend
on  inverted matrices requires  using the  covariances of  the inverse
matrix elements.   A concise formula  is developed in the  small error
limit for the covariance of  the inverted matrix elements. It is shown
to work  for nonsingular matrices. In particular  the determinant must
not become  zero when the matrix  elements are allowed  to vary within
their uncertainties.
 
An  attempt was  made  to study  how  well the  formula estimates  the
covariances for  a random  set of $2  \times2$ matrices  with positive
elements  between  zero and  one.   On  the  average the  formula  was
accurate to better than four percent with a root mean square deviation
equal to two percent for matrices that have determinants more than ten
sigmas from zero.
%

\section{Acknowledgements}
This project was supported by grants from the Natural Science and
Engineering Research Council of Canada.
%
%

%

\appendix              
\renewcommand{\theequation}{A-\arabic{equation}}
\setcounter{equation}{0}

\section{\large  \bf Derivation of the Covariance}

The formal solution to the system of real linear equations,
\begin{equation}
Y_{i} = A_{ij}X_{j} ~~~~i,j = 1\ldots,n
\end{equation}
where $X_{i}$ are the unknowns and  $A$ is a real $n \times n$ matrix,
is given by
%
\begin{equation}
X_{i} = A^{-1}_{ij}Y_{j}\:.
\end{equation}
A sum  over repeated indices  is assumed unless otherwise  noted.  The
uncertainties on the elements of $A$ are assumed to be known and given
by
\begin{equation}
\mathrm{cov}(A_{ij},A_{\alpha \beta})\:,
\end{equation}
the covariance of $A_{ij}$ and $A_{\alpha \beta}$.  If each element of
$A$  is   an  independent  random   quantity,  then  only   the  terms
$\mathrm{cov}(A_{ij},A_{ij})$  are  nonzero.   In  other  words,  each
element of the matrix will have an associated variance.

Similarly the uncertainties  on the elements of $Y$  are assumed to be
known  and  given  by  $\mathrm{cov}(Y_{i}, Y_{j})$.   Again,  if  the
elements of $Y$ are independent, the off-diagonal terms will be zero.

The most general form of the covariance for the $X_{i}$ is given by
the error propagation formula,
\begin{equation}
\mathrm{cov}(X_{i},X_{j}) = 
\frac{\partial X_{i}}{\partial A^{-1}_{\alpha \beta}}
\frac{\partial X_{j}}{\partial A^{-1}_{ab}}
\mathrm{cov} (A^{-1}_{\alpha \beta},A^{-1}_{ab}) 
+
\frac{\partial X_{i}}{\partial Y_{k}} 
\frac{\partial X_{j}}{\partial Y_{l}}
\mathrm{cov} (Y_{k}, Y_{l}) \:.
\label{eqtfullcov}
\end{equation}
The partial derivatives are given by:
\begin{eqnarray}
\frac{\partial X_{i}}{\partial A^{-1}_{\alpha \beta}} &
= &
\delta_{\alpha i}Y_{\beta} \:, \\
\frac{\partial X_{j}}{\partial A^{-1}_{a b}} &
= &
\delta_{aj}Y_{b} \:, \\
\frac{\partial X_{i}} {\partial Y_{k}} &
= &
A^{-1}_{ik} \:, \\
\frac{\partial X_{j}}{\partial Y_{l}} &
= &
A^{-1}_{jl} \:,
\end{eqnarray}
where $\delta_{ij}$ is the Kronecker delta.
Therefore, after relabelling summed-over indices, 
equation \ref{eqtfullcov} becomes
\begin{equation}
\mathrm{cov}(X_{i},X_{j})=Y_{\alpha}Y_{\beta}\mathrm{cov}
                            (A^{-1}_{i\alpha},A^{-1}_{j\beta})
+
A^{-1}_{ik}A^{-1}_{jl}\mathrm{cov}(Y_{k},Y_{l}) \:.
\end{equation}
The only unknown quantity is  
$\mathrm{cov}(A^{-1}_{i \alpha}, A^{-1}_{j\beta})$.

The inverse matrix elements can be considered as functions 
of the original matrix elements,
\begin{equation}
A^{-1}_{\alpha\beta} = A^{-1}_{\alpha\beta}(A_{ij})\:.
\end{equation}
The error propagation formula yields
\begin{equation}
\mathrm{cov} \left( 
A^{-1}_{\alpha\beta}, A^{-1}_{a b} 
\right) =
\frac{\partial A^{-1}_{\alpha \beta}} {\partial A_{ij}}
\frac{\partial A^{-1}_{ab}} {\partial A_{kl}}
\mathrm{cov} \left(
A_{ij}, A_{kl}
\right) \:. 
\label{covpartial}
\end{equation}
Hence the terms $\partial A^{-1}_{\alpha \beta} / 
\partial A_{ij}$
are required.

Consider the identity,
\begin{equation}
A^{-1}_{ij}A_{jk} = \delta_{ik}\:.
\end{equation}
Taking the derivative with respect to $A_{\alpha\beta}$ yields,
\begin{equation}
\frac {d\delta_{ik}} {dA_{\alpha \beta}} 
=  A^{-1}_{ij} 
\frac{\partial A_{jk}} {\partial A_{\alpha \beta}}
+ A_{jk}
\frac{\partial A^{-1}_{ij}} {\partial A^{-1}_{ab}}
\frac{\partial A^{-1}_{ab}} {\partial A_{\alpha \beta}} = 0\:.
\end{equation}
Making the substitutions,
\begin{equation}
\frac{\partial A_{jk}} {\partial A_{\alpha \beta}}
= \delta_{j \alpha} \delta_{k \beta}
\end{equation}
and
\begin{equation}
\frac{\partial A_{ij}^{-1}} {\partial A_{a b}^{-1}}
= \delta_{i a} \delta_{j b}
\end{equation}
yields
\begin{equation}
A^{-1}_{i\alpha} \delta_{k \beta} + A_{bk} 
\frac{\partial A^{-1}_{ib}} {\partial A_{\alpha\beta}} = 0\:.
\end{equation}
Contracting with $A^{-1}_{k \gamma}$ and using $A_{bk} A^{-1}_{k \gamma} =
\delta_{b \gamma}$ gives
\begin{equation}
\frac{\partial A^{-1}_{i\gamma}} {\partial A_{\alpha \beta}} 
= - A^{-1}_{i \alpha}A^{-1}_{\beta \gamma} \:,
\label{eqttensor}
\end{equation}
which can be relabelled as
\begin{equation}
\frac{\partial A^{-1}_{\alpha \beta}} {\partial A_{ij}}
= -A^{-1}_{\alpha i}A^{-1}_{j \beta} \:.
\end{equation}

Hence, equation \ref{covpartial} becomes
\begin{equation}
\mathrm{cov}(A^{-1}_{\alpha\beta},A^{-1}_{ab}) 
= A^{-1}_{\alpha i}A^{-1}_{j\beta}A^{-1}_{ak}
  A^{-1}_{lb}\mathrm{cov}(A_{ij},A_{kl}) \:.
\label{eqtfinal}
\end{equation}
This is the fundamental formula of this paper.

An important case is when matrix elements $A_{ij}$ are uncorrelated. 
If each element of $A_{ij}$ has associated with it a variance 
${[\sigma_{A}]}^{2}_{ij}$ then
\begin{equation}
\mathrm{cov}
\left(
A_{ij},A_{kl}
\right) = [\sigma_{A}]^{2}_{ij} \delta_{ik}\delta_{jl} \:
\: \: \mathrm{(no \ summation)}.
\label{eqtAdiag}
\end{equation}
Note that a corrected version of equation~\ref{eqtsimple} now
can be calculated as the 
variance $[\sigma_{A^{-1}}]^{2}_{\alpha \beta}$ for each element of 
$A^{-1}_{\alpha \beta}$,
\begin{equation}
[\sigma_{A^{-1}}]^{2}_{\alpha\beta}
\equiv
\mathrm{cov} 
(A^{-1}_{\alpha \beta}, A^{-1}_{\alpha \beta}) 
=
[A^{-1}]^{2}_{\alpha i} [\sigma_{A}]^{2}_{ij}
[A^{-1}]^{2}_{j \beta}\:.
\label {eqtdiagonal}
\end{equation}

Finally, the full covariance formula for uncorrelated $A_{ij}$ is 
given by
\begin{equation}
\mathrm{cov}(A^{-1}_{\alpha \beta} , A^{-1}_{a b}) =
(A^{-1}_{\alpha i} A^{-1}_{a i})
[\sigma_A]^2_{ij}
(A^{-1}_{j \beta} A^{-1}_{j b})
\label{eqtfull_cov_diag}
\end{equation}
where there is no sum inside the parentheses. For an $n\times n$ matrix
there are $n^4$ covariances. From symmetry of the covariance operation
one sees that there are
$n^2(n^2+1)/2$ independent covariance terms of 
which $n^2$ are diagonal, i.e. variances,
and $n^2(n^2-1)/2$ are ``off-diagonal''.

\renewcommand{\theequation}{B-\arabic{equation}}
\setcounter{equation}{0}
\section{\large  \bf An Explicit Two Dimensional Example}
The two dimensional case is very instructive because 
it is simple enough to derive analytically.
Consider
\begin{equation}
A = 
\left(
 \begin{array}{cc}
           a & b \\
           c & d 
  \end{array}
\right) \:,\:\: A^{-1} =
\left(
 \begin{array}{cc}
        \alpha & \beta\\
        \gamma &\delta
 \end{array}
\right)
\end{equation}
where
\begin{equation}
A^{-1} = \frac {1}{\mid A \mid} 
\left( 
  \begin{array}{cc}
            d & -b\\
           -c & a
  \end{array}
\right)
\label{eqtinverse2d}
\end{equation} 
and 
$\mid A \mid = ad-bc$ is the determinant of $A$.  The matrix elements of 
$A$ have uncertainties given by
\begin{equation}
\sigma_{A} =
\left(
   \begin{array}{cc}
          \sigma_{a} & \sigma_{b}\\
          \sigma_{c} & \sigma_{d}
   \end{array}
\right)\:.
\end{equation}
The uncertainty on $\alpha = A^{-1}_{11}$, for example, can be calculated using
the error propagation formula,
\begin{equation}
\sigma^{2}_{\alpha} = 
(\frac{\partial \alpha}{\partial a})^{2} \sigma_{a}^{2}
+
(\frac{\partial \alpha}{\partial b})^{2} \sigma_{b}^{2}
+
(\frac{\partial \alpha}{\partial c})^{2} \sigma_{c}^{2}
+
(\frac{\partial \alpha}{\partial d})^{2} \sigma_{d}^{2} \:.
\end{equation}
The partial derivatives can be calculated from equation~\ref{eqtinverse2d}
as for example,
\begin{equation}
\frac{\partial \alpha}{\partial a} =
\frac{\partial }{\partial a} \left (\frac{d}{ad-bc} \right ) =
\frac{\partial }{\partial a} \left (\frac{-d^2}{(ad-bc)^2} \right ) =
- \alpha^2 \:.
\end{equation}
For completeness the required derivatives are:
\begin{equation}
\frac{\partial \alpha}{\partial a} = -\alpha^{2} \:,\:\: 
\frac{\partial \alpha}{\partial b} = -\alpha \gamma \:,\:\:
\frac{\partial \alpha}{\partial c} = -\alpha \beta \:,\:\:
\frac{\partial \alpha}{\partial d} = -\beta \gamma \:,
\label{eqtpartials}
\end{equation}
and therefore
\begin{equation}
\sigma^{2}_{\alpha} = 
\alpha^{4}\sigma_{a}^{2} + \alpha^{2}\gamma^{2}\sigma_{b}^{2}
+
\alpha^{2}\beta^{2}\sigma_{c}^{2} + \beta^{2}\gamma^{2}\sigma_{d}^{2}\:.
\label{eqtexplicitvar}
\end{equation}
This result can be compared directly with the results 
from equation~\ref{eqtdiagonal},
where $\alpha = \beta = 1$,
\begin{equation}
\mathrm{cov} 
(A^{-1}_{1 1}, A^{-1}_{1 1})
\equiv 
[\sigma_{A^{-1}}]^{2}_{1 1} 
=
[A^{-1}]^{2}_{1 i} [\sigma_{A}]^{2}_{ij}
[A^{-1}]^{2}_{j 1}\:,
\end{equation}
which on substitution for the elements of $A^{-1}$ and $\sigma_{A}$
yields equation~\ref{eqtexplicitvar}.

Similarly an explicit example of an ``off-diagonal''  covariance
can be calculated,
\begin{equation}
\mathrm{cov}(\alpha , \beta ) =
\frac{\partial \alpha}{\partial a}
          \frac{\partial \beta}{\partial a} \sigma_{a}^{2}+
\frac{\partial \alpha}{\partial b}
           \frac{\partial \beta}{\partial b} \sigma_{b}^{2}+
\frac{\partial \alpha}{\partial c}
           \frac{\partial \beta}{\partial c} \sigma_{c}^{2}+
\frac{\partial \alpha}{\partial d}
           \frac{\partial \beta}{\partial d} \sigma_{d}^{2} \:,
\end{equation}
where the partial derivatives with respect to beta are given by,
\begin{equation}
\frac{\partial \beta}{\partial a} = -\alpha\beta \:,\:\: 
\frac{\partial \beta}{\partial b} = -\alpha \delta \:,\:\:
\frac{\partial \beta}{\partial c} = -\beta^2 \:,\:\:
\frac{\partial \beta}{\partial d} = -\beta \delta \:.
\end{equation}
Therefore the covariance of $\alpha$ and $\beta$ is given by,
\begin{equation}
\mathrm{cov}(\alpha,\beta) =
\alpha^3\beta\sigma_a^2 + 
\alpha^2\gamma\delta\sigma_b^2 +
\alpha\beta^3\sigma_c^2 +
\beta^3\gamma\delta\sigma_d^2 \:.
\end{equation}
This result is identical with equation~\ref{eqtfull_cov_diag}
where $\alpha=\beta=1$, $a=1$ and $b=2$. Note, that in general, 
the covariance is not zero.

\renewcommand{\theequation}{C-\arabic{equation}}
\setcounter{equation}{0}
\section{\large  \bf Covariances of the Determinant}
Let A be an $n \times n$ matrix with elements $A_{ij}$. The determinant
is given by,
\begin{equation}
   \det A = |A| = \epsilon_{\alpha_1 \alpha_2 \dots \alpha_n}
A_{1\alpha_1}A_{2\alpha_2} \dots A_{n\alpha_n} \:,
\end{equation}
where $\epsilon_{\alpha_1 \alpha_2 \dots \alpha_n}=1$ for cyclic
permutations of $\alpha_i$, a factor of -1 is applied each
time any two indices are exchanged, hence 
$\epsilon_{\alpha_1 \alpha_2 \dots \alpha_n}$ is zero if any two indices
are the same. This can be rewritten by defining new matrices
by removing from $A$ the elements of its $i^{th}$ row and $j^{th}$
column. The determinant of the remaining $(n-1) \times (n-1)$
matrix is called~\cite{ref_mathbook} the minor of $A_{ij}$. If the
minor of $A_{ij}$ is denoted by $M_{ij}$ then the cofactor of
$A_{ij}$ is given by,
\begin{equation}
  C_{ij} = (-1)^{i+j} M_{ij}.
\end{equation}
The matrix of all cofactors is $C$. The determinant
of $A$ can now be written as,
\begin{equation}
\det A = A_{ij}C_{ij} = A_{ji}C_{ji},  
\label{eqt_det}
\end{equation}
for $i$ fixed.

The error on $\det A$ is given by,
\begin{equation}
\sigma^2_{|A|} = \frac{\partial \det A}{\partial A_{ab} }
   \frac{\partial \det A}{\partial A_{\alpha\beta} }
   \mathrm{cov}(A_{ab},A_{\alpha\beta})
\end{equation}
From equation~\ref{eqt_det} one gets,
\begin{equation}
    \frac{\partial \det A}{\partial A_{ab} } = C_{ab},
\end{equation}
since $i=a$ is fixed and $C_{ij}$ is independent of
$A_{ab}$ for all $j$.
Therefore, one obtains the very compact result,
\begin{equation}
  \sigma^2_{|A|} = C_{ab}C_{\alpha\beta}  
        \mathrm{cov}(A_{ab},A_{\alpha\beta}).
                                              \label{eqt_deterr}
\end{equation}
\end{document}